\documentstyle[aps,prl,preprint,epsf]{revtex}
\input psfig

\newcounter{saveeqn}
\newcommand{\alphaeqn}{\setcounter{saveeqn}{\value{equation}}%
\stepcounter{saveeqn}\setcounter{equation}{0}%
\renewcommand{\theequation}{\mbox{\arabic{saveeqn}\alph{equation}}}}
\newcommand{\reseteqn}{\setcounter{equation}{\value{saveeqn}}%
\renewcommand{\theequation}{\arabic{equation}}}

\begin{document}

\title{Fast and slow dynamics of hydrogen bonds in liquid water}

\author{Francis~W. Starr$^1$, Johannes
K. Nielsen$^{1,2}$ \& H.~Eugene Stanley$^1$}

\address{$^1$Center for Polymer Studies, Center for Computational
Science, and Department of Physics, Boston University, Boston, MA
02215 USA}

\address{$^2$Department of Mathematics and Physics, Roskilde University,
\\ Postbox 260, DK-4000 Roskilde, Denmark}

\date{September 22, 1998}
 
\maketitle

\begin{abstract}
We study hydrogen-bond dynamics in liquid water at low temperatures
using molecular dynamics simulations.  We find that bond lifetime
(``fast dynamics'') has Arrhenius temperature dependence.  We also
calculate the bond correlation function and find that the correlation
time (``slow dynamics'') shows power-law behavior.  This power-law
behavior, as well as the decay of the bond correlations, is consistent
with the predictions of the mode-coupling theory.  The correlation time
at the lowest temperature studied shows deviation from power-law
behavior that suggests continuity of dynamic functions between the
liquid and glassy states of water at low pressure.
\end{abstract}
\bigskip
\pacs{PACS numbers: 64.60.My, 64.70.Ja, 64.70.Pf}

Both experiments~\cite{debenedetti,chen-teixeira,tex-book,mc-book} and
simulations~\cite{sf89,spsh90,lc96} of water have focused on
understanding various aspects of hydrogen bond dynamics, such as the
average bond lifetime $\tau_{HB}$ and the structural relaxation time
$\tau_R$.  Experiments show that $\tau_{HB}$ follows a simple Arrhenius
law, but that characteristic relaxation times commonly obey power laws
on supercooling~\cite{debenedetti,chen-teixeira}.  Simulations are
particularly useful for investigating water in the supercooled regime
since nucleation does not occur on the time scale of the simulations and
access to quantitative hydrogen bond information is available.
Accordingly, we carry out simulations of normal and supercooled water
and find Arrhenius behavior of $\tau_{HB}$, as expected from
experimental results~\cite{chen-teixeira,dls-experiments}; the
activation energy associated with $\tau_{HB}$ provides a simple test of
the criteria used in the simulations to define a bond.  Further, we find
power-law behavior of $\tau_R$ which breaks down at the lowest
temperature.  This non-trivial growth of $\tau_R$ is consistent with a
continuous transition of the supercooled liquid to the glassy state.

To elucidate how simple temperature dependence of $\tau_{HB}$
[Fig.~\ref{lifetime-relax}] can lead to complex dependence of $\tau_R$
[Fig.~\ref{relaxation-scaling}], we perform lengthy molecular-dynamics
simulations (up to 70~ns, over 7 orders of magnitude) at seven
temperatures between 200~K and 350~K~\cite{simulations}.  We use the
extended simple point charge (SPC/E) potential~\cite{spce} for water and
calculate the bond dynamics by considering two definitions of an intact
hydrogen bond: (i) an {\it energetic} definition~\cite{sf89}, which
considers two molecules to be bonded if their oxygen-oxygen separation
is less than 3.5 \AA~and their interaction energy is less than a
threshold energy $E_{HB}$, and (ii) a {\it geometric}
definition~\cite{lc96}, which uses the same distance criterion but no
energetic condition, instead requiring that the O--H...O angle between
two molecules must be less than a threshold angle $\theta_{HB}$.

We calculate $\tau_{HB}$ using both bond definitions over the entire
temperature range simulated and find Arrhenius behavior of $\tau_{HB}$
[Fig.~\ref{lifetime-relax}].  Measurements of $\tau_{HB}$ using
depolarized light scattering
techniques~\cite{chen-teixeira,dls-experiments} find Arrhenius
behavior~\cite{path}.  Previous simulations did not find Arrhenius
behavior of $\tau_{HB}$ \cite{spsh90}.  The activation energy $E_A$
associated with $\tau_{HB}$ has been interpreted as the energy required
to break a bond via librational motion, a ``fast''
motion~\cite{chen-teixeira,dls-experiments}.  Comparison of experimental
and simulated values of $E_A$ provides a primitive test of the bonding
criteria in our simulations; we obtain reasonable agreement between
experimental and simulated values of $E_A$ using thresholds of $E_{HB} =
-10$ kJ/mol for the energetic definition and $\theta_{HB} = 30
^\circ$~\cite{lc96} for the geometric definition.  We find better
quantitative agreement with experiments for $\tau_{HB}$ values obtained
from the geometric definition than for $\tau_{HB}$ values obtained from
energetic definition -- possibly because the geometric bond definition,
like the depolarized light scattering experiments, is highly sensitive
to the linearity of the bond.  We also calculate $\tau_{HB}$ using the
thresholds $E_{HB} = 0$~kJ/mol~\cite{sf89,spsh90} and $\theta_{HB} =
35^\circ$ and find Arrhenius behavior, but with $E_A$ roughly 30\%
smaller for the energetic definition, and roughly 10\% smaller for the
geometric definition.

The quantity $\tau_{HB}$ is one characteristic time -- the mean -- of
the distribution of bond lifetimes $P(t)$, which measures the
probability that an initially bonded pair remains bonded at all times up
to time $t$, and breaks at $t$.  $P(t)$ is obtained from simulations by
building a histogram of the bond lifetimes for each configuration, and
may be related to $\tau_{HB}$ by $\tau_{HB} = \int_0^\infty t P(t)
dt$~\cite{pt-note}.  For both bond definitions and all seven
temperatures simulated, we calculate the behavior of
$P(t)$~[Fig.~\ref{lifetimes}(a)].  We observe neither power-law nor
exponential behavior for either bond definition, unlike previous
calculations using the ST2 potential which revealed power-law dependence
of $P(t)$~\cite{spsh90}.  The difference in $P(t)$ is not surprising,
since the previous study considered a different threshold value and also
followed a path near a liquid-liquid critical
point~\cite{pses92,spce-cp}.  The behavior of $P(t)$ for the two bond
definitions is different, likely caused by differences in sensitivity of
the definitions to librational motion.

We study $\tau_R$ (``slow'' dynamics) for an initially bonded pair by
calculating the bond correlation function $c(t)$, the probability that a
randomly chosen pair of molecules is bonded at time $t$ provided that
the pair was bonded at $t=0$ (independent of possible breaking in the
interim time). To calculate $c(t)$, we define

\begin{equation}
c(t) \equiv {\langle h(0)h(t) \rangle}/{\langle h^2 \rangle },
\label{correlation-function}
\end{equation}
 
\noindent where $h(t)$ is a binary function for each pair of molecules
$\{i,j\}$, and $h(t)=1$ if molecules $\{i,j\}$ are bonded at time $t$
and $h(t)=0$ if $\{i,j\}$ are not bonded at time
$t$~\cite{lc96}.  The angular brackets denote an average of
all pairs $\{i,j\}$ and starting times.  We define $\tau_R$ by
$c(\tau_R) \equiv e^{-1}$. Short time fluctuations and choice of bond
definition do not strongly affect the long-time behavior of the
``history-independent'' quantity $c(t)$ (which {\it does not\/} depend
on the continuous presence of a bond), but such fluctuations cause both
qualitative and quantitative differences in the long-time behavior of
the ``history-dependent'' $P(t)$ (which {\it does\/} depend on the
continuous presence of a bond).

We can interpret the behavior of $\tau_R$ in terms of the mode-coupling
theory (MCT) for a supercooled liquid approaching a glass
transition~\cite{debenedetti,mct1}.  In accordance with MCT, we find --
independent of the two bond definitions considered -- power-law growth
for $T \ge 210$~K

\alphaeqn
\begin{equation}
\tau_R \sim (T-T_c)^{-\gamma},
\label{plaw}
\end{equation}

\noindent with $\gamma = 2.7 \pm 0.1$ and $T_c = 197.5 \pm 1.0$~K,
approximately 50~K less than the temperature of maximum density $T_{MD}$
of SPC/E [Fig.~\ref{relaxation-scaling}(a)]~\cite{tc-gamma-note}.  Fits
of experimental relaxation times to Eq.~(\ref{plaw}) also find $T_c$ at
a temperature 50~K less than the $T_{MD}$ of water~\cite{sa76}.  In MCT,
$T_c$ is the temperature of the ideal kinetic glass transition and is
larger than the glass transition temperature $T_g$ determined from,
e.g., viscosity or relaxation time measurements.  At $T = 200$~K,
$\tau_R$ is smaller than would be estimated by Eq.~(\protect\ref{plaw}),
most likely because MCT does not account for activated processes which
aid diffusion and reduce relaxation times at low
temperatures~\cite{not-crazy}.  Typically, these activated processes
become important near $T_c$, as observed here.

Our simulation results for $\tau_R$ can also be fit by the
Vogel-Fulcher-Tammann (VFT) form~\cite{debenedetti} for $T \ge 210$~K

\begin{equation}
\tau_R \sim e^{A/(T-T_0)},
\label{vft}
\end{equation}
\reseteqn

\noindent with $T_0 = 160$~K [Fig.~\ref{relaxation-scaling}(b)].  In the
entropy theory of the glass transition
\protect\cite{debenedetti,adam-gibbs}, $T_0$ is associated with the
Kauzmann temperature, the temperature where the extrapolated entropy of
the supercooled liquid approaches the entropy of the solid.  For typical
liquids we expect $T_0<T_g$, so estimating $T_0$ and $T_c$ provides
lower and upper bounds for $T_g$.  For water, however, fits to
Eq.~(\ref{vft}) of experimental data (which are far above $T_g$) yield
$T_0>T_g$ \cite{debenedetti}; hence we do not consider our $T_0$ value
to be a lower bound for $T_g$ of the SPC/E model.  $T_g$ is defined
experimentally as the temperature where the viscosity reaches $10^{12}$
Pa$\cdot$s or $\tau = 100$~s.  Experiments near $T_g$ often show a
crossover from VFT behavior to ``normal'' Arrhenius
behavior~\cite{debenedetti}.  While our simulations are still relatively
far from $T_g$ (based on the value of $\tau_R$), a naive extrapolation
assuming that temperatures $T=210$~K and $200$~K follow Arrhenius
behavior yields $T_g \approx 105$~K~\cite{pg}.  Applying the same shift
to $T_g$ as is observed for the $T_{MD}$ of SPC/E relative to water
(specifically 35~K), the speculated $T_g$ of SPC/E is consistent with
experimental measurements of $T_g \approx 140$~K in
water~\cite{debenedetti}.

A continuous crossover to Arrhenius behavior in water might account for
the fact that $T_g < T_0$ when experimental data are fit to
Eq.~(\ref{vft}).  VFT or power law behavior (``fragile'') for $T \gtrsim
220$~K changing to Arrhenius behavior (``strong'') for $T \lesssim
220$~K could smoothly connect the structural relaxation times in the
liquid with those of the glass.  A fragile-to-strong transition in water
has been previously suggested \cite{angell93}, and recent experimental
results for the diffusion constant (which scales as $\tau_R^{-1}$) at
temperatures closer to $T_g$ may help to determine if such a transition
occurs ~\cite{smith-kay}.

The reactive flux, defined by the derivative 

\alphaeqn
\begin{equation}
k(t) \equiv -{d c(t)}/{d t},
\end{equation}

\noindent measures the effective decay rate of an initial set of
hydrogen bonds.  At temperature $T=300$~K, non-exponential decay of
$k(t)$ was found for the closely-related SPC model using the geometric
bond definition~\cite{lc96}.  Our calculations of $k(t)$
[Fig.~\ref{lifetimes}(b)] reveal a power-law region $k(t) \sim
t^{-\zeta}$ for $T \ge 250$~K for both bond definitions, with an
exponent $\zeta = 0.5 \pm 0.1$.  The range of the power-law region
increases from about one decade at 350~K to about two decades at 250~K.

The value of $\zeta$ can be interpreted using MCT, which predicts that
$c(t)$ decays from a plateau value $c_p$ with power-law dependence

\begin{equation}
c_p - c(t) \sim t^b
\label{plateau}
\end{equation}
\reseteqn

\noindent in the range where $k(t)$ appears to be
power-law~\cite{debenedetti,mct1}.  From Eq.~(\ref{plateau}), $k(t) \sim
t^{b-1}$, so $b = 1 - \zeta$.  We find $b=0.5 \pm 0.1$, consistent with
previous work~\cite{sgtc-96}, further suggesting that the bond behavior
is consistent with MCT predictions for a glass transition.  For T $<
250$~K, the decay of $k(t)$ appears to be neither power-law nor
exponential.

We note that even though the functional form of $\tau_{HB}$ does not
appear to be strongly dependent on bond definition, $P(t)$ is different
for the two definitions---suggesting that $P(t)$ may not be the best
function for studying bond dynamics.  In contrast $k(t)$---which
includes bond reformation---appears to be largely independent of the
bond definition at long times~\cite{still75}.  

The {\it Arrhenius} behavior and short time scale of $\tau_{HB}$
indicates that librational motion is a thermally activated process that
is largely independent of the structural slowing.  The {\it
non-Arrhenius} behavior of $\tau_R$ and the non-exponential relaxation
of $k(t)$ are consistent with the presence of the proposed ideal kinetic
glass transition for SPC/E approximately 50~K below the $T_{MD}$ of
SPC/E water~\cite{sgtc-96}, which coincides with the temperature at
which many experimental relaxation times appear to diverge~\cite{sa76}.
It was hypothesized that the apparent singular temperature of liquid
water observed experimentally may be identified with the $T_c$ of
MCT~\cite{sgtc-96,plsl87}; our results support this hypothesis.  They
also complement scenarios that account for the anomalous thermodynamic
behavior, such as the liquid-liquid transition
hypothesis~\cite{pses92,roberts,mc98} or the the singularity-free
hypothesis~\cite{sing-free}.

We thank C.A.~Angell, S.V.~Buldyrev, S.-H.~Chen, I.~Gro{\ss}e,
S.T.~Harrington, A.~Luzar, O.~Mishima, P.~Ray, F.~Sciortino,
A.~Skibinsky, D.~Stauffer, and especially S.~Havlin and S.~Sastry for
helpful discussions and/or comments on the manuscript.  All simulations
were performed using the Boston University 192-processor SGI/Cray Origin
2000 supercomputer.  FWS is supported by a NSF graduate fellowship.  The
Center for Polymer Studies is supported by NSF grant CH9728854 and
British Petroleum.

\newbox\figa
\setbox\figa=\psfig{figure=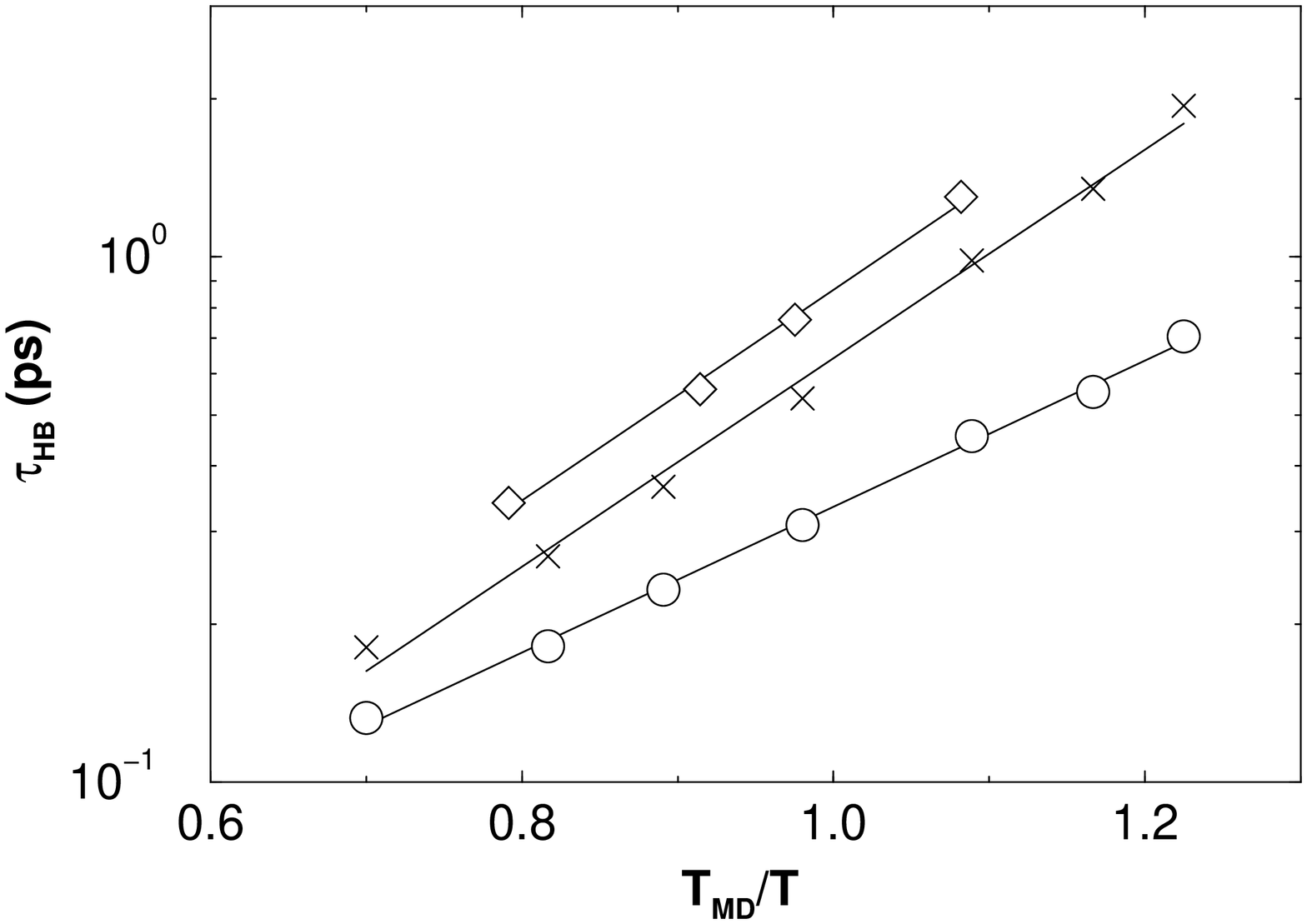,width=15cm,angle=-90}
\begin{figure*}[htbp]
\begin{center}	
\leavevmode	
\centerline{\box\figa}
\narrowtext
\caption{Average bond lifetime $\tau_{HB}$ for the energetic ($\circ$)
and geometric ($\times$) bond definitions; shown for comparison are
experimental data ($\diamond$) for depolarized light
scattering~\protect\cite{chen-teixeira}.  We observe that all of the
data can be fit by Arrhenius behavior $\tau_{HB} = \tau_0 \exp(E_A/kT)$
with approximate activation energies: (i) $E_A=8.8 \pm 1.0$~kJ/mol
(energetic definition); (ii) $E_A=9.3 \pm 1.0$~kJ/mol (geometric
definition); (iii) $E_A=10.8 \pm 1.0$~kJ/mol (experimental).  We scale
the temperature of the simulation results by $T_{MD}^{SPC/E}$ and
temperature of the experimental data by $T_{MD}^{H_2O}$ to facilitate
comparison with with results~\protect\cite{bc94,hpss97,bbk97}.}
\label{lifetime-relax}
\end{center}
\end{figure*}

\newbox\figa
\setbox\figa=\psfig{figure=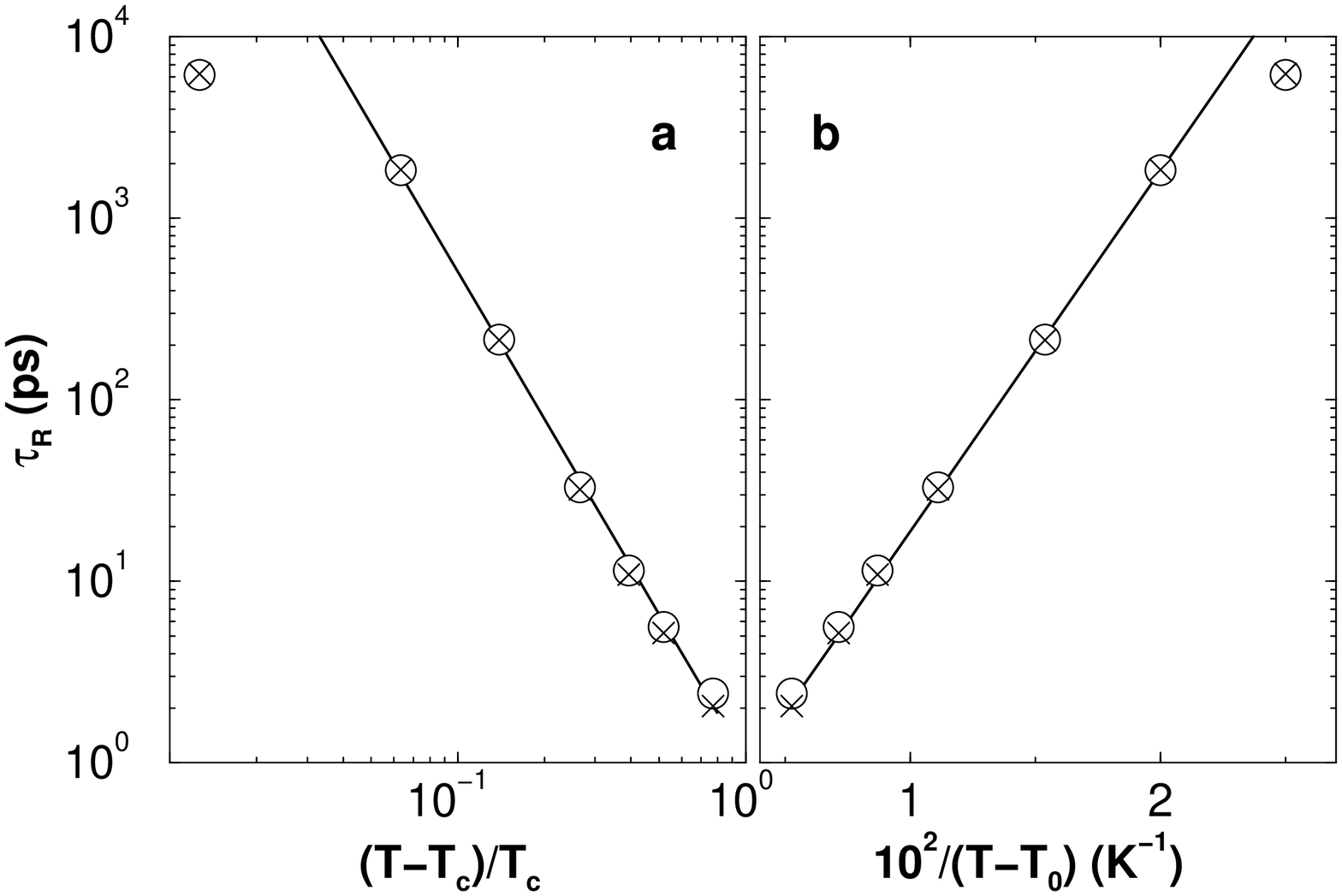,width=15cm,angle=-90}
\begin{figure*}[htbp]
\begin{center}	
\leavevmode	
\centerline{\box\figa}
\caption{Relaxation time $\tau_R$ of the hydrogen bond correlation
function $c(t)$ for the energetic ($\circ$) and the geometric ($\times$)
bond definitions. (a) Fit to the scaling form predicted by mode-coupling
theory (solid line) with $T_c = 197.5$~K. In MCT, $T_c$ is the
temperature of structural arrest.  (b) Fit to the VFT form (red line)
with $T_0 = 160$~K.  The deviation from both fitting forms we consider
may indicate a smooth transition relaxations time in supercooled water
with that of glassy water. }
\label{relaxation-scaling}
\end{center}
\end{figure*}

\newbox\figa
\setbox\figa=\psfig{figure=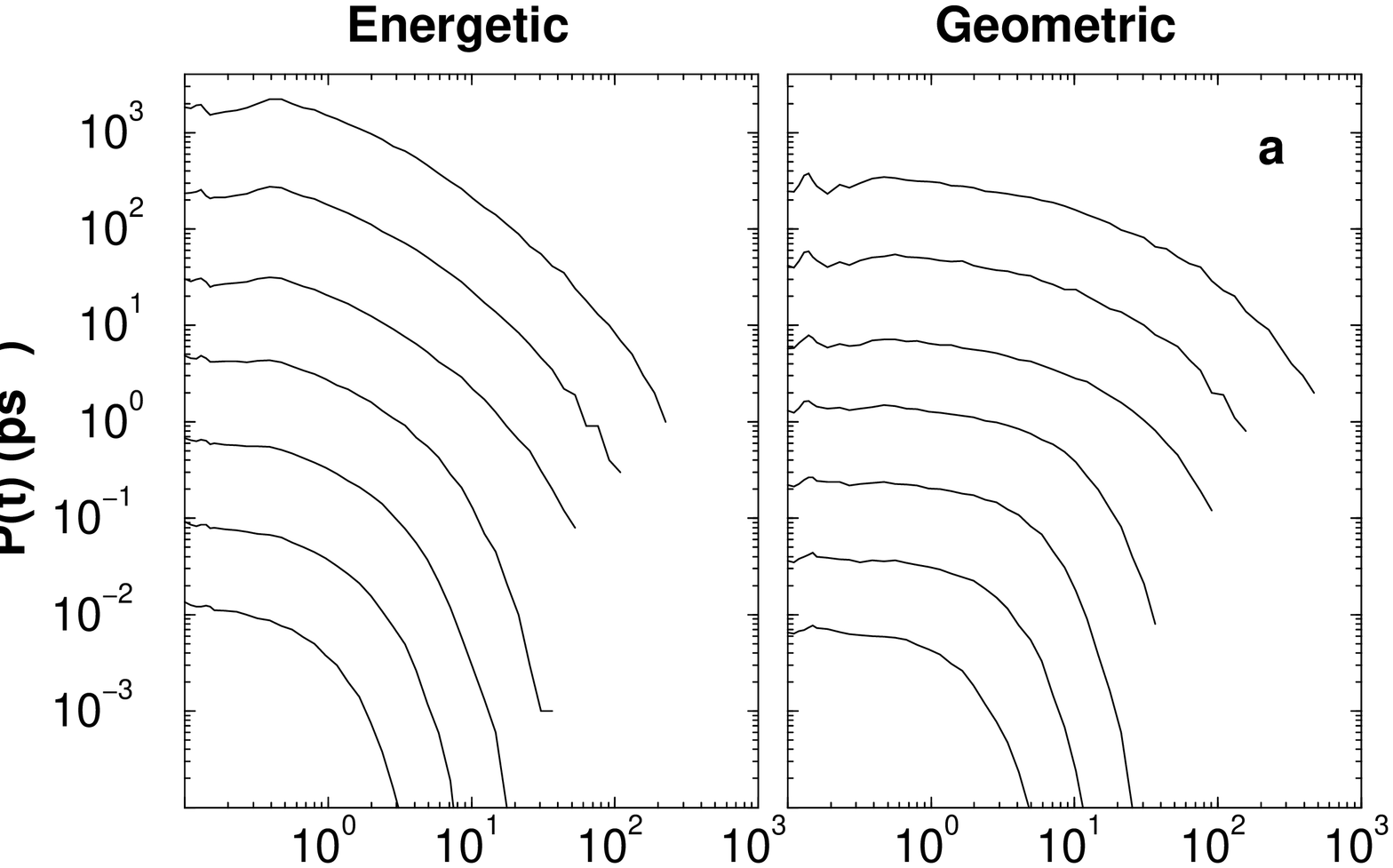,width=15cm,angle=-90}
\newbox\figb
\setbox\figb=\psfig{figure=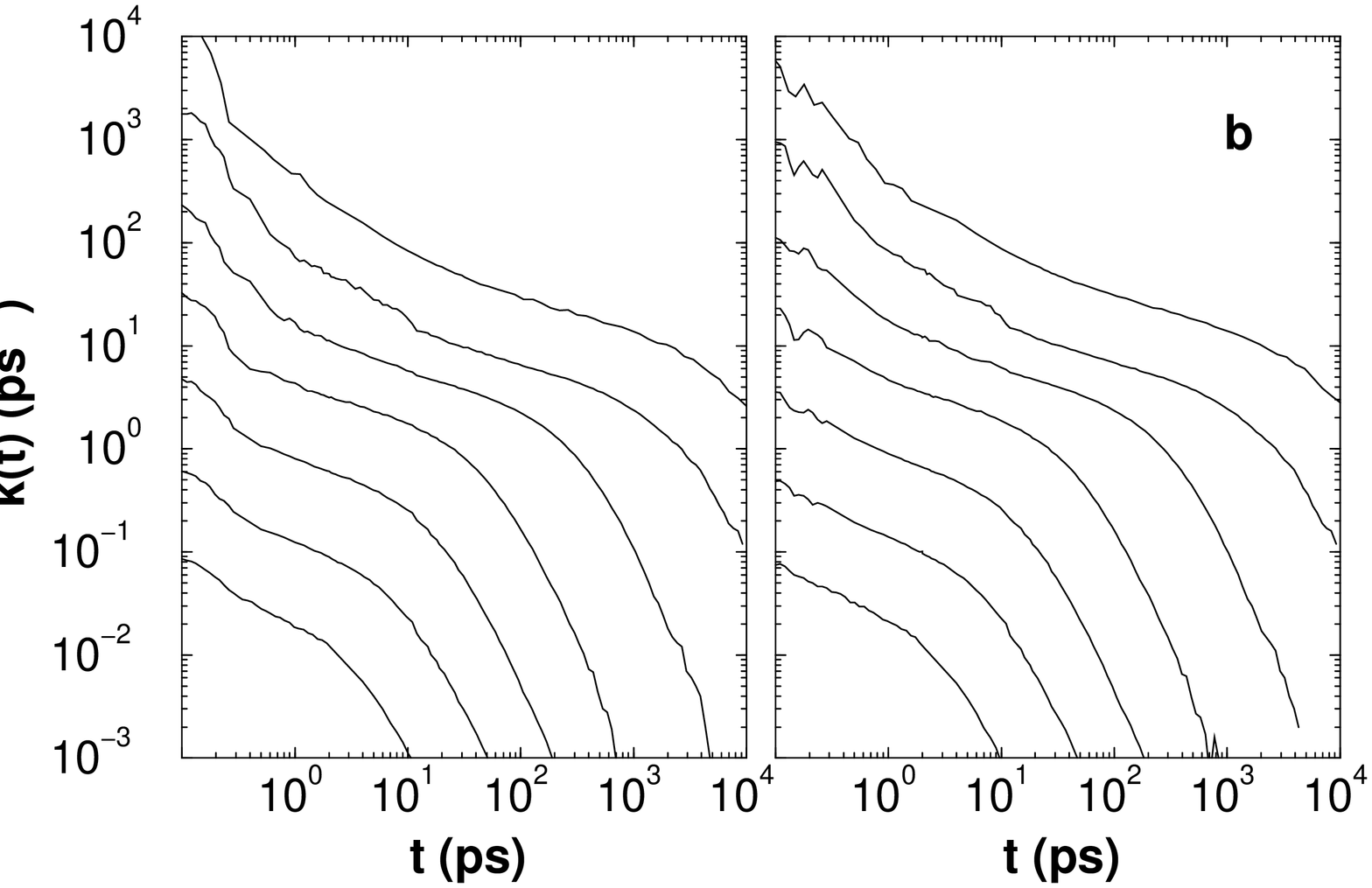,width=15cm,angle=-90}
\begin{figure*}[htbp]
\begin{center}	
\leavevmode	
\centerline{\box\figa}
\centerline{\box\figb}
\narrowtext
\caption{Time-dependent relaxation of the hydrogen bonds.  (a) The
bond lifetime distribution $P(t)$.  The curves can be identified as
follows (reading from top to bottom): 200~K, 210~K, 225~K, 250~K, 275~K,
300~K, and 350~K.  Each curve is offset by one decade for clarity.  (b) The reactive flux $k(t)$. Note that $k(t)$ does not depend on the
unbroken presence of a bond, so $k(t)$ decays less rapidly than $P(t)$.
After a transient period of rapid librational motion up to $t\approx
0.3$~ps, we observe a region of power-law decay for $T$ is above 250~K.
For $t \protect\gtrsim 0.3$~ps, $k(t)$ is nearly identical for both bond
definitions, a result which likely arises because both definitions use
the same distance criterion.  We calculate $k(t)$ from the numerical
derivative of $c(t)$, which is well-approximated by a stretched
exponential for $t\protect\gtrsim 0.3$~ps.  We average over all pairs in
the system and many initial starting times for trajectories ranging in
length from 200~ps at 350~K, to 40~ns at 200~K.  Note that our results
for the geometric definition at $T=300$~K are consistent with recent
calculations for the SPC potential (see inset of Fig.~1 of
ref.~\protect\cite{lc96}).}
\label{lifetimes}
\end{center}
\end{figure*}


\begin{thebibliography}{10}
\bibitem{path} 
The experiments of refs.~\cite{chen-teixeira,dls-experiments}
were performed at constant pressure.  The simulations we present are at
constant density, thus a quantitative comparison on the lifetime values.
However, the change in pressure along the 1.0~g/cm$^3$ is relatively
small~\cite{hpss97}, and we still expect Arrhenius behavior.
\bibitem{pt-note}
Since $P(t)$ depends on the unbroken presence of a bond, $P(t)$ is
sensitive to the sampling frequency.  Thus choosing a long time interval
between sampled configurations corresponds to ignoring processes where a
bond is broken for a short time and subsequently reforms.  To calculate
$P(t)$, we sample every 10 fs (shorter than the typical libration time,
which could destroy a bond).  We also measured $P(t)$ with a sampling
frequency of 1~fs and observed no noticeable change in the decay of
$P(t)$.
\bibitem{spce-cp} If a liquid-liquid transition occurs in SPC/E, it has
been estimated to terminate at a critical point located at 160~K and
200~MPa~\cite{hpss97}.  The present simulations should not be
significantly affected by this critical point, as we are far from it.
\bibitem{tc-gamma-note}
The values $T_c$ and $\gamma$ depend on the pressure.  Studying
different dynamic properties, ref.~\cite{sgtc-96} independently
calculated $T_c$ and $\gamma$ for the same potential and -- depending on
the pressure considered -- obtained ($T_c=186$~K, $\gamma=2.3$) and
($T_c=199$~K, $\gamma=2.8$).  Thus the bond dynamics appear to follow
the same phenomenology as the commonly observed dynamic quantities.
\end{thebibliography}
\end{document}